\documentclass[a4paper,11pt]{article}
\pdfoutput=1 

\usepackage{jcappub} 

\usepackage[utf8]{inputenc}
\usepackage[english]{babel}
\usepackage[T1]{fontenc}
\usepackage{bbm}
\usepackage[margin=2.5cm]{geometry}               
\usepackage{amssymb}
\usepackage{amsmath}	
\newcommand\numberthis{\addtocounter{equation}{1}\tag{\theequation}}

\usepackage{graphicx}	
\usepackage{wrapfig}
\usepackage{braket}
\usepackage{stackengine,scalerel}
\def\tang{\ThisStyle{\abovebaseline[0pt]{\scalebox{-1}{$\SavedStyle\perp$}}}}

\title{\boldmath Disforming to Conformal Symmetry}

\author[a]{Pavel Jirou\v{s}ek,}
\author[c]{Keigo Shimada,}
\author[b]{Alexander Vikman}
\author[c]{and \qquad Masahide Yamaguchi}

\affiliation[a]{\it High Energy Physics, Cosmology \& Astrophysics Theory Group, University of Cape Town,\\Private Bag, Cape Town 7700, South Africa}
\affiliation[b]{CEICO -{\it Central European Institute for Cosmology and Fundamental Physics},\\ FZU -{\it Institute of Physics of the Czech Academy of Sciences,\\Na Slovance 1999/2, 18221 Prague 8, Czech Republic}}
\affiliation[c]{\it Department of Physics, Tokyo Institute of Technology,\\Tokyo 152-8551, Japan}

\emailAdd{pavel.jirousek@uct.ac.za}
\emailAdd{shimada.k.ah@m.titech.ac.jp}
\emailAdd{vikman@fzu.cz}
\emailAdd{gucci@phys.titech.ac.jp}

\abstract{We analyse the dynamical properties of disformally transformed theories of gravity. We show that disformal transformation typically introduces novel degrees of freedom, equivalent to the mimetic dark matter, which possesses a Weyl-invariant formulation. We demonstrate that this phenomenon occurs in a wider variety of disformal transformations than previously thought.}

\begin{document}
\maketitle
\flushbottom
\section{Introduction}
Ongoing observational advances in physics of cosmological inflation, dark energy and dark matter resulted in a revival of interest in theories where gravity is modified on astrophysical and cosmological scales. Arguably the simplest theories of this type should involve only one additional dynamical degree of freedom on top of the usual gravitons. This minimalistic restriction on the number of the extra degrees of freedom is manifestly realized in the Horndeski theories \cite{Horndeski:1974,Deffayet:2011gz,Kobayashi:2011nu} where all equations of motion are second order. In this way the additional dynamical degree of freedom appears explicitly in the action as a scalar field variable $\phi$. However, the true degrees of freedom can sometimes be hidden and can reveal themselves only after a tedious Hamiltonian/Dirac analysis of a constrained system. In particular, a degenerate system with higher-derivative equations of motion can still have only one extra degree of freedom. This idea is realized in the "beyond Horndeski" or (GLPV) \cite{Gleyzes:2014dya,Gleyzes:2014qga} theories in and degenerate higher-order scalar-tensor
theories (DHOST) theories \cite{Langlois:2015cwa}, for recent reviews see e.g. \cite{Kobayashi:2019hrl,Langlois:2018dxi}. It turned out, some of these degenerate theories can be obtained from the Horndeski construction by introducing a non-minimal coupling of the scalar field to matter, see \cite{Zumalacarregui:2013pma}. Namely, this non-minimal coupling appears through an \emph{invertible} disformal transformation \cite{Bekenstein:1992pj} of the metric 
\begin{equation}
g_{\mu\nu}=C\left(Y,\phi\right)h_{\mu\nu}+D\left(Y,\phi\right)\partial_{\mu}\phi\,\partial_{\nu}\phi\,,\label{eq:disformal transformation}
\end{equation}
where we denoted
\begin{equation}
Y=h^{\mu\nu}\partial_{\mu}\phi\,\partial_{\nu}\phi\,.\label{eq:Y}
\end{equation}
Consequently, the Horndeski part of the action only depends on $h_{\mu\nu}$ and $\phi$, while matter lives in the spacetime given by the disformally transformed metric $g_{\mu\nu}$ defined by (\ref{eq:disformal transformation}). For instance, a dust-like matter moves along the timelike geodesics of $g_{\mu\nu}$ instead of those of $h_{\mu\nu}$. If the transformation (\ref{eq:disformal transformation}) is invertible, one can express $h_{\mu\nu}$ in terms of $g_{\mu\nu}$ and $\phi$ and then substitute this expression into the Horndeski action. In this way, matter becomes minimally coupled, while the resulting theory has higher order equations of motion due to the presence of the derivatives in (\ref{eq:disformal transformation}).

Crucially, the dynamical content of the minimally coupled theory is equivalent to that of the original non-minimally coupled Horndeski model. It follows then that the number of degrees of freedom remains unchanged despite the presence of higher derivatives in the action. This way we can easily generate higher order scalar-tensor theories beyond Horndeski (more precisely of the class DHOST Ia) without running into the risk of introducing additional degrees of freedom to the model, \cite{Bettoni:2013diz,Domenech:2015tca,Crisostomi:2016czh,BenAchour:2016cay,Crisostomi:2016tcp,Takahashi:2017pje}, for exceptions from this rule see \cite{Deffayet:2020ypa}. 
In particular, this construction of the new theories through the disformal transformations provides a convenient method to investigate compact objects in these theories, for recent studies see e.g. \cite{BenAchour:2020wiw,Babichev:2020qpr,Anson:2020trg,BenAchour:2020fgy,Minamitsuji:2020jvf,Anson:2021yli,Faraoni:2021gdl,Babichev:2022awg}.
The caveat here is that the stated equivalence and our ability to switch between the two descriptions (frames) relies heavily on the invertibility of the transformation, which is not guaranteed for a generic choice of functions $C$ and $D$. Correspondingly, non-invertible transformations have been shown to introduce additional degrees of freedom to the theory and thus they change the dynamical content. A notable class of such non-invertible transformations has been studied in \cite{Deruelle:2014zza}. This class is defined by the following relation between the functions $C\left(Y,\phi\right)$ and $D\left(Y,\phi\right)$
\begin{equation}
    D(Y,\phi)=-\frac{C(Y,\phi)}{Y}+c(\phi)\ .\label{eq:regularity condition natalie}
\end{equation}
It has been demonstrated in \cite{Deruelle:2014zza} that applying these transformations to General Relativity (GR) with usual minimally coupled matter results in an emergence of a novel dynamical sector equivalent to the mimetic gravity \cite{Chamseddine:2013kea}, for review see e.g. \cite{Sebastiani:2016ras}. In the latter construction, the appearance of the novel hidden degree of freedom is connected to gauge degeneracy in form of the Weyl symmetry. 
\\
\\
Clearly, it is crucial to understand which disformal transformations are "safe" - are just changes of description, and which secretly introduce novel dynamical degrees of freedom. In this paper, we address this issue. We consider a general {\it{seed}} theory consisting of usual matter and a general metric theory of gravity including GR. We show that transformation with a typical choice of functions $C\left(Y,\phi\right)$
and $D\left(Y,\phi\right)$ violating relation \eqref{eq:regularity condition natalie} still does introduce novel degrees of freedom when applied to such a seed theory. Namely, this happens on solutions $\phi_{\star}(x)$ of the first order nonlinear partial differential equation (PDE)
\begin{equation}
    C=C_{Y}Y+D_{Y}Y^{2}\ ,\label{eq:regularity condition zuma}
\end{equation}
were the subscript denotes a partial derivative (e.g. $C_{Y}\equiv\partial C/\partial Y$). Note that functions $C\left(Y,\phi\right)$
and $D\left(Y,\phi\right)$ related by \eqref{eq:regularity condition natalie} automatically satisfy equation \eqref{eq:regularity condition zuma}, so that generic configurations $\phi(x)$ satisfy this PDE. Clearly, there are many more functions $C\left(Y,\phi\right)$
and $D\left(Y,\phi\right)$ which allow for existence of a solution of \eqref{eq:regularity condition zuma} then those related through \eqref{eq:regularity condition natalie}. The additional matter sector emerging on these configurations $\phi_{\star}(x)$ is either equivalent to the original mimetic dark matter (DM) \cite{Chamseddine:2013kea} or to its analytical or tachyonic continuation to spacelike gradients of $\phi$. We will refer to both cases under the name {\it{mimetic gravity}}. In turn, mimetic DM is classically equivalent to irrotational dust \cite{Golovnev:2013jxa,Barvinsky:2013mea,Hammer:2015pcx}. Mimetic DM can well describe cosmology on large/linear scales. Our analysis shows that mimetic gravity is much more common in disformally transformed theories than previously thought. \par
This paper is organized as follows. In section \ref{Jacobi}, following and extending results of \cite{Zumalacarregui:2013pma}, we study Jacobian of the general disformal transformation \eqref{eq:disformal transformation} and present condition for appearance of the new degree of freedom \eqref{eq:regularity condition zuma}. In section \ref{Tensor Equations} we vary the disformed action with $h_{\mu\nu}$ and show that configurations $\phi_{\star}(x)$ satisfying equation \eqref{eq:regularity condition zuma} result in emergence of the energy-momentum tensor corresponding to mimetic gravity in the seed tensor equation of motion e.g. in the Einstein equation in case of GR. Thus, the tensor equation of motion does not involve $h_{\mu\nu}$. In section \ref{Scalar EoM} we vary the disformed action with respect to field $\phi$. We show that this equation of motion takes the form of covariant continuity or conservation equation  written again in terms of disformed $g_{\mu\nu}$. In section \ref{Examples} we illustrate our findings with simple examples. In particular we point out that one can construct such functions $C\left(Y,\phi\right)$ and $D\left(Y,\phi\right)$ that it is impossible to satisfy condition \eqref{eq:regularity condition zuma}. However, most choices of these functions will result in some solutions of \eqref{eq:regularity condition zuma}. In section \ref{Mimetic Weyl} we derive an equivalent action for conformal ($D=0$) subclass of transformations \eqref{eq:disformal transformation}. We show that this action corresponds to mimetic gravity. Based on these results, in section \ref{Action_Disformal} we reduce disformed action to the one effectively equivalent to that of mimetic gravity. 
Finally, we discuss our findings and propose future direction of research in section \ref{Conclusions}.

\section{The Jacobian: Singularity and the Hamilton-Jacobi equation}\label{Jacobi}

Before we inspect the dynamical consequences of a general disformal transformation $g_{\mu\nu}\left(h_{\sigma\rho},\phi\right)$ given by \eqref{eq:disformal transformation} let us first revisit the properties of its Jacobian matrix. This matrix can be easily calculated since the transformation does not depend on the derivatives of the metric $h_{\mu\nu}$. Hence, we get
\begin{equation}
    \mathcal{J}_{\mu\nu}^{\rho\sigma}=\frac{\partial g_{\mu\nu}}{\partial h_{\rho\sigma}}=\frac{1}{2}C\left(\delta_{\mu}^{\sigma}\delta_{\nu}^{\rho}+\delta_{\nu}^{\sigma}\delta_{\mu}^{\rho}\right)-\left(C_{Y}h_{\mu\nu}+D_{Y}\phi_{,\mu}\phi_{,\nu}\right)\phi^{,\sigma}\phi^{,\rho} ,\label{eq:Jacobian matrix}
\end{equation}
where the upper indexes of the partial derivatives ($\phi_{,\mu}\equiv\partial_{\mu}\phi$) on the right hand side have been raised using the contravariant metric $h^{\mu\nu}$, so that
\begin{equation}
\phi^{,\mu}\equiv h^{\mu\nu}\phi_{,\nu}\ .
\end{equation}
On the other hand, we use notation  
\begin{equation}
\nabla^{\mu}\phi\equiv g^{\mu\nu}\phi_{,\nu}\ ,
\end{equation}
where $\nabla_{\mu}$ is the Levi-Civita connection preserving the metric $g_{\mu\nu}$. We consider $\mathcal{J}^{\sigma\rho}_{\mu\nu}$ as a $10$ by $10$ matrix whose "indexes" are parameterized by a symmetric pair of spacetime indexes. In the above equation the upper "index" is the pair $\sigma\rho$ while the lower "index" is the pair $\mu\nu$. Much of the properties of this matrix can be glanced from its eigenvalues and their associated eigenvectors. These are obtained by solving the following equation
\begin{equation}
    \mathcal{J}^{\mu\nu}_{\sigma\rho}\,\xi^{a}_{\mu\nu}=\lambda_{a}\,\xi^{a}_{\sigma\rho}\ ,\label{eq:eigenvector equation zuma}
\end{equation}
where $\xi^{a}_{\mu\nu}$ is the eigenvector (symmetric eigentensor) associated to $\lambda_{a}$. The index $a$ labels all possible solutions of this equation and it is not summed over here. In \cite{Zumalacarregui:2013pma} it was showed that there are only two distinct eigenvalues
\begin{align}
    \lambda_{0}=&C\ ,\nonumber\\
    \lambda_{\star}=&C-C_{Y}Y-D_{Y}Y^{2}\ ,\label{eq:eigenvalues zuma}
\end{align}
for general disformal transformation \eqref{eq:disformal transformation}. The determinant of $\mathcal{J}^{\mu\nu}_{\sigma\rho}$ is the product of all the eigenvalues raised to the power of their respective multiplicities. Thus, as long as {\it{all}} eigenvalues, $\lambda_a$, are non-vanishing one has $\text{det}\mathcal{J}_{\mu\nu}^{\rho\sigma}\neq0$. Consequently, the mapping $g_{\mu\nu}\left(h_{\sigma\rho},\phi \right)$ is locally invertible with respect to $h_{\mu\nu}$ by virtue of the Inverse Function Theorem. We will refer to transformations with vanishing determinant of the Jacobian as \emph{singular disformal transformations}. \par
 
If we further require that $g_{\mu\nu}$ constitutes a valid metric tensor we find that the eigenvalue $\lambda_{0}$ cannot vanish. Indeed, the calculation of the tensor inverse to $g_{\mu\nu}$ straightforwardly gives
\begin{equation}
    g^{\mu\nu}=\frac{1}{C}\left ( h^{\mu\nu}-\frac{D}{C+DY}\phi^{,\mu}\phi^{,\nu}\right)\ ,\label{eq:inverse metric}
\end{equation}
where again $\phi^{,\mu}\equiv h^{\mu\nu}\phi_{,\nu}$. This inverse tensor is clearly ill defined when $\lambda_{0}=C=0$. The non-existence of $g^{\mu\nu}$ would be clearly incompatible with $g_{\mu\nu}$ being a metric tensor. Therefore, the vanishing of the determinant is necessarily tied to 
\begin{equation}
\lambda_{\star}\left(\phi,Y\right)=C-C_{Y}Y-D_{Y}Y^{2}=0 ,\label{eq:singularity condition}
\end{equation}
in any physically viable scenarios with singular disformal transformation\footnote{
Requiring that $g_{\mu\nu}$ is a good metric tensor implies further restrictions on the functions $C$ and $D$ that go beyond eigenvalue problem. For example, from the inverse metric above we can additionally see that $C+DY\neq 0$. In particular, this implies that $c(\phi)\neq 0$ in degeneracy condition \eqref{eq:regularity condition natalie}.  Additionally posing $C+DY>0$ ensures that the two metrics both have a Lorentzian signature. These requirements and others have been discussed in \cite{Bekenstein:1992pj,Bettoni:2013diz,Bruneton:2006gf}. 
}. 
The implicit function theorem provides algebraic solution  $Y_\star(\phi)$ of equation \eqref{eq:singularity condition}, then the equation of motion for the scalar field  
\begin{equation}
h^{\mu\nu}\partial_{\mu}\phi\,\partial_{\nu}\phi=Y_\star(\phi)\,,
\label{eq:Hamilton_Jacobi}
\end{equation}
is of the type of the relativistic Hamilton-Jacobi equation \cite{Landafshitz_Teorpol}.
Expression for the contravariant disformed metric \eqref{eq:inverse metric} yields the relation
\begin{equation}
X\equiv g^{\mu\nu}\phi_{,\mu}\phi_{\nu}=\frac{Y}{C+DY}\ ,
\label{eq:X(Y)}
\end{equation}
useful for the forthcoming discussion. In particular, differentiating this relation provides
\begin{equation}
\partial_{Y}X=\frac{\lambda_{\star}}{\left(C+DY\right)^{2}}\ .
\label{eq:dXdY}
\end{equation}
Substituting in equation \eqref{eq:X(Y)} solution $Y_\star(\phi)$ one disforms the Hamilton-Jacobi equation to 
\begin{equation}
g^{\mu\nu}\partial_{\mu}\phi\,\partial_{\nu}\phi=X_{\star}(\phi)\ .
\label{eq:EoM_for_phi_in_g}
\end{equation}
These equations always have a solution. 
Equation \eqref{eq:singularity condition} can in general admit multiple solutions $Y^i_\star(\phi)$ labeled with an index $i$. For each of these branches  there is a corresponding solution of the Hamilton-Jacobi equation $\phi^i_\star(x)$. 
Interestingly, various solutions obtained this way can differ in character, as the functions $Y^i_\star(\phi)$ can be positive or negative. Therefore, both, timelike and spacelike gradients of $\phi$ are admissible. Formally, it is also possible to obtain solutions with a light-like gradients; however, for functions $C$ and $D$ sufficiently smooth in $Y$, these configurations unavoidably imply $C=0$, due to the singularity condition \eqref{eq:singularity condition}. As we have argued earlier, this is not compatible with $g_{\mu\nu}$ being a metric tensor.
Thus, the timelike or spacelike character of the solutions should be preserved by evolution, i.e. $X_{\star}(\phi)$ should not change its sign. On the other hand the different branches of solutions may glue together or one solution can branch out, when the condition of the inverse function theorem is not satisfied so that 
\begin{equation}
C_{YY}+2D_{Y}+D_{YY}Y=0\,.
\end{equation}
Contrary to purely conformal transformations, here the timelike gradients of $\phi$ with respect to the metric $h_{\mu\nu}$ can, in principle, become spacelike with respect to the metric $g_{\mu\nu}$, and vice versa. Applying \eqref{eq:X(Y)}
we can see that as long as $C+DY>0$ the character of solutions agree in both metrics. This assumption is often taken to ensure that the metric $g_{\mu\nu}$ has a proper Lorentzian signature, see e.g. \cite{Bettoni:2013diz}. 

Solving equation \eqref{eq:eigenvector equation zuma} provides the eigenvectors
\begin{align}
    \xi^{0}_{\mu\nu}&=\phi^{\perp}_{\mu\nu}\ ,\nonumber\\
    \xi^{\star}_{\mu\nu}&=C_{Y}h_{\mu\nu}+D_{Y}\phi_{,\mu}\phi_{,\nu}\ ,\label{eq:eigenvectors zuma}
\end{align}
which play a central role in our argument later. Here $\phi^{\perp}_{\mu\nu}$ is any symmetric tensor that is orthogonal to $\phi^{,\mu}\phi^{,\nu}$, i.e.
\begin{equation}
    \phi^{\perp}_{\mu\nu}\,\,\phi^{,\mu}\phi^{,\nu}=0\ .\label{eq:perpendicularity zuma}
\end{equation}
The set of eigenvectors \eqref{eq:eigenvectors zuma} allows us to determine the multiplicities of the associated eigenevalues. Indeed, the condition \eqref{eq:perpendicularity zuma} singles out 9 independent covariant symmetric tensors, which then constitute the eigenspace associated to $\lambda_{0}$. Hence the multiplicity of $\lambda_{0}$ is 9. On the other hand the eigenspace associated with $\lambda_{\star}$ is clearly one dimensional and it is generated only by $\xi^{\star}_{\mu\nu}$. Hence, the multiplicity of $\lambda_{\star}$ is one. Consequently, we can be sure that we have found all the eigenvectors, as the associated vector space is 10 dimensional.
\\
\\
Interestingly, for vanishing $\lambda_{\star}$ the eigenvector $\xi^{\star}_{\mu\nu}$  can be interpreted as a generator of a symmetry transformation. Indeed, an infinitesimal change of the metric
\begin{equation}
    \delta_{\epsilon} h_{\mu\nu}=\epsilon\,\xi^{\star}_{\mu\nu}\ ,
    \label{eq:Weyl_generator}
\end{equation}
where $\epsilon$ is a small parameter, generates the corresponding transformation of $g_{\mu\nu}$
\begin{equation}
    \delta_{\epsilon} g_{\mu\nu}=\epsilon\,\mathcal{J}^{\sigma\rho}_{\mu\nu}\,\xi^{\star}_{\sigma\rho}=\epsilon\,\lambda_{\star}\,\xi^{\star}_{\mu\nu}\ .
\end{equation}
This clearly vanishes when $\lambda_{\star}=0$ and thus \eqref{eq:Weyl_generator} becomes a symmetry of the theory. It is important to stress that it is not guaranteed that this infinitesimal transformation can in general be integrated to a continuous transformation. However, in simple examples we can find the corresponding continuous transformation. For instance, the original mimetic dark matter \cite{Chamseddine:2013kea} corresponds to $C=Y$ and $D=0$. Infinitesimal transformation \eqref{eq:Weyl_generator} then reduces to
\begin{equation}
    \delta_{\epsilon} h_{\mu\nu}=\epsilon\,\xi^{\star}_{\mu\nu}=\epsilon\,h_{\mu\nu}\ ,
\end{equation}
which is clearly the generator of the Weyl transformations
\begin{equation}
    h_{\mu\nu}\rightarrow e^{2\omega}h_{\mu\nu}\ .
\end{equation}
Here $\omega(x)$ parameterizes the transformation and does not need to be infinitesimal. As another example, one can consider the disformal transformation with $C=1$ and $D=1-1/Y$ which results  
\begin{equation}
    \delta_{\epsilon} h_{\mu\nu}=\epsilon\,\xi^{\star}_{\mu\nu}=\epsilon'\,\partial_{\mu}\phi\partial_{\nu}\phi\ ,
\end{equation}
where in the last equality we redefined the infinitesimal parameter $\epsilon$ to absorb $Y^{-2}$. This corresponds to additive shifts by $\partial_{\mu}\phi\partial_{\nu}\phi$
\begin{equation}
    h_{\mu\nu}\rightarrow h_{\mu\nu}+\omega \, \partial_{\mu}\phi\partial_{\nu}\phi\ .
\end{equation}\par
 While the set of eigenvectors \eqref{eq:eigenvectors zuma} is complete, it is possible to obtain an additional set of {\it dual} or {\it left} eigenvectors, $\zeta_a^{\mu\nu}$, satisfying
 \begin{equation}
    \zeta_a^{\sigma\rho}\,\mathcal{J}^{\mu\nu}_{\sigma\rho}=\lambda_a\,\zeta_a^{\mu\nu}\ ,\label{eq:eigenvector equation dual}
\end{equation}
instead of \eqref{eq:eigenvector equation zuma}. Here the associated eigenvalues, $\lambda_a$, are identical to those from \eqref{eq:eigenvalues zuma}. 
The solutions of the above equation provide the following dual eigeinvectors
\begin{align}
    \zeta_{0}^{\mu\nu}&=\phi_{\tang}^{\mu\nu}\ ,\nonumber\\
    \zeta_{\star}^{\mu\nu}&=\phi^{,\mu}\phi^{,\nu}\ ,\label{eq:eigenvectors dual}
\end{align}
where $\phi_{\tang}^{\mu\nu}$ is any symmetric tesnor that is perpendicular to $\xi^{\star}_{\mu\nu}$, i.e.
\begin{equation}
    \zeta_{0}^{\mu\nu}\xi^{\star}_{\mu\nu}=\phi_{\tang}^{\mu\nu}\xi^{\star}_{\mu\nu}=0\ .
\end{equation}
Note that equation \eqref{eq:perpendicularity zuma} can be rewritten in a similar manner using $\zeta_{\star}^{\mu\nu}$ as
\begin{equation}
    \phi^{\perp}_{\mu\nu}\zeta_{\star}^{\mu\nu}=\xi^{0}_{\mu\nu}\zeta_{\star}^{\mu\nu}=0\ .\label{eq:perpendicularity dual}
\end{equation}
Similarly to $\xi^{\star}_{\mu\nu}$ the dual eigenvector $\zeta^{\mu\nu}_{\star}$ becomes physically significant for the singular disformal transformations. As we will demonstrate in the upcoming section, such transformations give rise to an additional degree of freedom equipped with an energy-momentum tensor proportional to $\zeta^{\mu\nu}_{\star}$.



\section{Metric Dynamics due to Singular Disformal Transformation}\label{Tensor Equations}
Dynamical consequences of disformal transformations \eqref{eq:disformal transformation} performed simultaneously both in the Einstein-Hilbert action and in the matter action have been explored in \cite{Deruelle:2014zza}. There it has been demonstrated that disformal transformations satisfying condition \eqref{eq:regularity condition natalie} introduce an additional matter sector to the theory. 
In this section we show that these dynamics appear even when the weaker condition \eqref{eq:regularity condition zuma} is satisfied as a differential equation. Our analysis is thus more general than \cite{Deruelle:2014zza}, as \eqref{eq:regularity condition natalie} is not necessarily valid in such cases.  We demonstrate this for a general theory that is obtained by a disformal transformation from an arbitrary \emph{seed} action $S_{\mathrm{seed}}[g,\Psi_{M}]$ 
\begin{equation}
    S_{\mathrm{dis}}[h,\phi,\Psi_{M}]=S_{\mathrm{seed}}[g(h,\phi),\Psi_{M}]\ ,\label{eq:action disformed}
\end{equation}
where $\Psi_{M}$ denote matter fields which, however, play no role in the following argument. For simplicity, we will assume that the seed action does not depend on $\phi$, contrary to many interesting mimetic theories and equivalents e.g. \cite{Chamseddine:2014vna,Mirzagholi:2014ifa,Capela:2014xta,Lim:2010yk,Arroja:2015wpa,Ramazanov:2015pha}. 
Variation of disformed action \eqref{eq:action disformed} with respect to the independent metric $h_{\mu\nu}$ yields tensor equation of motion 
\begin{equation}
    \frac{\delta S_{\mathrm{dis}}}{\delta h_{\mu\nu}}=\frac{\delta S_{\mathrm{seed}}}{\delta g_{\sigma\rho}}\,\mathcal{J}^{\mu\nu}_{\sigma\rho}=0\ .\label{eq:general disformal EoM}
\end{equation}
The properties of solutions of this equation significantly depend on whether the Jacobian $\mathcal{J}^{\mu\nu}_{\sigma\rho}$ is a regular or a singular matrix. 

In the regular case, the Jacobian can be inverted and the resulting equations of motion match those of the original seed theory when they are interpreted as equations for $g_{\mu\nu}$:
\begin{equation}
    \frac{\delta S_{\mathrm{seed}}}{\delta g_{\mu\nu}}=0\ .\label{eq:general EoM}
\end{equation}

In the singular case, the Jacobian matrix $\mathcal{J}^{\mu\nu}_{\sigma\rho}$ has a non-trivial kernel, which, for physically relevant solutions, is generated by the left eigenvector, $\zeta_{\star}^{\sigma\rho}$, from \eqref{eq:eigenvectors dual}  corresponding to the vanishing $\lambda_{\star}$. Thus, equation of motion \eqref{eq:general disformal EoM} implies that $\delta S_{\mathrm{seed}}/\delta g_{\sigma\rho}$ is an element of this kernel. In other words, equation \eqref{eq:general EoM} receives a non-trivial right hand side proportional to this eigenvector: 
\begin{equation}
   \frac{\delta S_{\mathrm{seed}}}{\delta g_{\mu\nu}}=\bar{\rho}\, \zeta_{\star}^{\mu\nu}=\bar{\rho}\,\phi^{,\mu}\phi^{,\nu}\ .\label{eq:general disformal EoM 2}
\end{equation}

The proportionality factor $\bar{\rho}(x)$ introduces a single undetermined density function, which signals the appearance of novel degree of freedom in the theory. Interestingly, the novel term is identical for all singular disformal transformations. Furthermore, the term clearly corresponds to the energy-momentum tensor for the mimetic gravity. This link becomes even more clear when we consider usual gravity minimally interacting with matter fields $\Psi_{M}$. The seed theory is described by the sum of the   Einstein-Hilbert action and an action for matter. Then, variation with respect to the metric $g_{\mu\nu}$ yields\footnote{We use the reduced Planck units $8\pi G_N=1$ and signature convention $\left(+,-,-,-\right)$.} standard result 
\begin{equation}
   \frac{2}{\sqrt{-g}}\, \frac{\delta S_{\mathrm{seed}}}{\delta g_{\mu\nu}}=G^{\mu\nu}-T^{\mu\nu}\ ,
   \label{standard_metric_variation}
\end{equation}
where $T^{\mu\nu}$ is the energy-momentum tensor of $\Psi_{M}$. Consequently, equation of motion \eqref{eq:general disformal EoM 2} becomes
\begin{equation}
    G^{\mu\nu}-T^{\mu\nu}=\frac{\rho}{Y\left(C+DY\right)}\phi^{,\mu}\phi^{,\nu}\ ,\label{eq:GR disformal EoM}
\end{equation}
where we introduced a convenient rescaling for the proportionality factor $\bar{\rho}$
\begin{equation}
\rho\equiv\frac{2}{\sqrt{-g}}\,g_{\mu\nu}\,\frac{\delta S_{\text{seed}}}{\delta g_{\mu\nu}}=\frac{2}{\sqrt{-g}}\,\bar{\rho}\,Y\,\left(C+DY\right) \ .
\label{eq:density}
\end{equation}
It is instructive to lower indices in \eqref{eq:GR disformal EoM} using the metric $g_{\mu\nu}$
\begin{equation}
G_{\mu\nu}-T_{\mu\nu}=\rho\,\frac{\left(C+DY\right)}{Y}\phi_{,\mu}\phi_{,\nu}=\rho\,\frac{\phi_{,\mu}\phi_{,\nu}}{X}\ ,
\label{eq:Enstein_low_index}
\end{equation}
where in the last equality we used equation \eqref{eq:X(Y)}. 
Hence, for timelike gradients (with respect to $g_{\mu\nu}$) singular disformal transformation results in just adding fluid-like dust to the seed theory. This dust has energy density $\rho$ and four velocity
\footnote{if $\partial_{\mu}\phi$ is not future-, but past-directed one should write minus in front of this expression.}
\begin{equation}
u_{\mu}=\frac{\partial_{\mu}\phi}{\sqrt{X_{\star}}}\ ,
\label{eq:four_velocity}
\end{equation}
where $X_{\star}\left(\phi\right)$ is obtained from $Y_{\star}\left(\phi\right)$ via relation \eqref{eq:X(Y)}. 
As it is well known, see e.g. \cite{Lim:2010yk}, such four-velocities describe timelike geodesics - have vanishing acceleration\footnote{For spacelike gradients one should change $X_{\star}\rightarrow -X_{\star}$ in expression \eqref{eq:four_velocity}, so that equation \eqref{eq:no_a} would describe spacelike geodesics.} 
\begin{equation}
u^{\lambda}\nabla_{\lambda}u^{\mu}=0\ .
\label{eq:no_a}
\end{equation}
Equation \eqref{eq:Enstein_low_index} becomes identical to the Einstein equation in mimetic gravity after field redefinition 
\begin{equation}
d\Phi=\frac{d\phi}{\sqrt{\left|X_{\star}\left(\phi\right)\right|}}\ ,
\label{eq:PHI}
\end{equation}
so that for timelike configurations, $\Phi$ is the four-velocity potential 
\begin{equation}
u_{\mu}=\partial_{\mu}\Phi\ .
\label{eq:potential}
\end{equation}
Clearly, for other seed theories the left hand side of the equation \eqref{eq:Enstein_low_index} should be changed following equation \eqref{standard_metric_variation}. \par
The existence of a non-trivial kernel of the Jacobian matrix further manifests itself as an extra symmetry of the equations of motion. By inspecting the original set of equations \eqref{eq:general disformal EoM} we can see that there is an invariance under the shift
\begin{equation}
    \frac{\delta S_{\mathrm{seed}}}{\delta g_{\mu\nu}}\rightarrow \frac{\delta S_{\mathrm{seed}}}{\delta g_{\mu\nu}}+\bar{\gamma}\,\zeta^{\mu\nu}_{\star}\ .\label{eq:EoM shift}
\end{equation}
The $\bar{\alpha}$ here is an arbitrary spacetime density. For the Einstein-Hilbert seed action this can be understood \cite{Jirousek:2022vdq} as a shift of either the energy momentum-tensor of $\Psi_{M}$ or of the Einstein tensor itself 
\begin{equation}
    G^{\mu\nu}\rightarrow G^{\mu\nu}+\gamma_{1}\zeta^{\mu\nu}_{\star}\ ,\qquad\mathrm{or}\qquad T^{\mu\nu}\rightarrow T^{\mu\nu}+\gamma_{2}\zeta^{\mu\nu}_{\star}\ .
\end{equation}
Here $\gamma_{1,2}$ are arbitrary scalar functions. The symmetry can be found even in the equation \eqref{eq:general disformal EoM 2}, where the above transformation can be compensated by a corresponding shift of the scalar $\rho$.



It is worthy to note that the appearance of the new term on the right hand side of the equation \eqref{eq:general disformal EoM 2} is reminiscent of the appearance of novel terms due to restricted variation. Restricted variation is often used as a possible way to obtain unimodular gravity \cite{Einstein:1919gv,Henneaux:1989zc,Buchmuller:1988wx,Buchmuller:1988yn}. In the context of the  unimodular gravity the Einstein-Hilbert action is varied while enforcing an additional condition on the variation of the metric 
\begin{equation}
    g^{\mu\nu}\tilde{\delta}g_{\mu\nu}=0\ .\label{eq:unimodular_restriction}
\end{equation}
This introduces a novel term $\varepsilon\, g^{\mu\nu}$ in the Einstein equation and results in the dynamics of unimodular gravity, see e.g. \cite{Unruh:1988in,Henneaux:1989zc, Alvarez:2006uu}. Similarly, the entire term $\mathcal{J}^{\mu\nu}_{\sigma\rho}\delta h_{\mu\nu}$ can be understood as a restricted variation
\begin{equation}
    \tilde{\delta}g_{\mu\nu}=\mathcal{J}^{\mu\nu}_{\sigma\rho}\delta h_{\mu\nu}\ ,
\end{equation}
which, in contrast to the standard variation $\delta g_{\mu\nu}$, satisfies the condition
\begin{equation}
    \zeta^{\mu\nu}_{\star}\tilde{\delta}g_{\mu\nu}=0\ ,
\end{equation}
in the singular case. This introduces a novel term $\bar{\rho}\, \zeta^{\mu\nu}_{\star}$ in the tensor equations of motion. Similar type of restricted variation can be used to obtain global dynamics of the Newton or even of the Planck constants \cite{Jirousek:2020vhy}. This raises an interesting question whether all types of such restricted variations of the metric can be realized through some metric transformation. For instance, unimodular restriction on variation of the metric \eqref{eq:unimodular_restriction} has been realized through different mimetic transformations involving vector fields \cite{Jirousek:2018ago,Hammer:2020dqp}.




\section{Scalar Dynamics due to Singular Disformal Transformation}\label{Scalar EoM}
Let us now inspect the role of the field $\phi$ in the two regimes of the theory, that is, when the Jacobian is either regular or singular. 
By assumption, the action $S_\text{dis}$ defined through \eqref{eq:action disformed} depends on $\phi$ through the metric $g_{\mu\nu}$ only. The chain rule applied to variation with respect to $\phi$ yields 
\begin{equation}
    \frac{\delta S_{\text{dis}}}{\delta\phi\left(x\right)}=\int d^{4}y\,\frac{\delta S_{\text{seed}}}{\delta g_{\sigma\rho}\left(y\right)}\,\frac{\delta g_{\sigma\rho}\left(y\right)}{\delta\phi\left(x\right)}\ .\label{chain_rule}
\end{equation}
Hence, equation of motion for $\phi$ is trivially satisfied as a consequence of tensor equation of motion \eqref{eq:general disformal EoM}, which in the regular case results in \eqref{eq:general EoM}.
In this regime $\phi$ is an unphysical gauge field, which we are free to chose. However, as we have discussed earlier, there are requirements on the configurations of $\phi$, which ensure that both $h_{\mu\nu}$ and $g_{\mu\nu}$ are proper metric tensors. So we subject our choice of $\phi$ to these conditions. 

In the singular regime instead of equation \eqref{eq:general EoM} one has 
\eqref{eq:general disformal EoM 2}, so that \eqref{chain_rule} provides a nontrivial differential equation.
The key observation of this paper is that when \eqref{eq:regularity condition natalie} is not satisfied, the factor $\lambda_{\star}$ appears as an overall factor of one of the equations of motion. Thus, it is natural to interpret singularity condition \eqref{eq:singularity condition} as a differential equation for $\phi$ determining its  dynamics. We can see this by contracting tensor equation of motion \eqref{eq:general disformal EoM} with the eigenvector $\xi^{\star}_{\mu\nu}$, which gives 

\begin{equation}
    \frac{\delta S_{\mathrm{seed}}}{\delta g_{\mu\nu}}\,\xi^{\star}_{\mu\nu}\,\,\lambda_{\star}=0\ .\label{eq:contraction with xi}
\end{equation}
The left hand side is clearly a product of two factors and the equation is satisfied if either of them is zero. Hence, we see that any $\phi$ satisfying $\lambda_{\star}=0$ is a solution of \eqref{eq:contraction with xi}. 

Since $\phi$ is obtained from \eqref{eq:Hamilton_Jacobi} the variation of the disformed action  with respect to $\phi$ allows us to determine $\rho$ defined in equation \eqref{eq:density} for a general seed theory. The variation gives 
\begin{equation}
\frac{\delta S_{\text{seed}}}{\delta g_{\mu\nu}}\,\frac{\partial g_{\mu\nu}}{\partial\phi}=2\partial_{\alpha}\left[\left(\frac{\delta S_{\text{seed}}}{\delta g_{\mu\nu}}\,\frac{\partial g_{\mu\nu}}{\partial Y}\,h^{\alpha\beta}+\frac{\delta S_{\text{seed}}}{\delta g_{\alpha\beta}}D\right)\phi_{,\beta}\right]\ ,
\label{eq:delta phi variation}
\end{equation}
which using equations \eqref{eq:density} and \eqref{eq:general disformal EoM 2} or \eqref{eq:Enstein_low_index}, after straight calculations, results in 
\begin{equation}
\nabla_{\mu}\left(\rho\,\left[C_{Y}+D_{Y}Y+D\right]\,\nabla^{\mu}\phi\right)=\frac{1}{2}\,\rho\,\left[\frac{C_{\phi}+D_{\phi}Y}{C+DY}\right]\ .
\label{eq:div rho intermid}
\end{equation}
Using singularity condition \eqref{eq:singularity condition} and its solution ${X}_{\star}\left(\phi\right)$ one can rewrite above equation as 
\begin{equation}
\nabla_{\mu}\left[\frac{\rho}{X_{\star}}\,\nabla^{\mu}\phi\right]=-\frac{1}{2}\,\rho\,\left[\frac{X_{\phi}}{X}\right]_{\star}\ .
\label{eq:div rho source}
\end{equation}
Utilizing relation \eqref{eq:dXdY} and definition \eqref{eq:PHI} this equation can be further simplified as
\begin{equation}
\nabla_{\mu}\left[\rho\,\nabla^{\mu}\Phi\right]=0\ ,
\label{eq:div rho universal}
\end{equation}
valid for timelike and spacelike gradients. For timelike gradients, 
one can utilize the definition of four-velocity \eqref{eq:four_velocity} and write this equation in form of the conservation equation 
\begin{equation}
\nabla_{\mu}\left(\rho\,u^{\mu}\right)=0\ .
\label{eq:continuity}
\end{equation}


Notably, the above equations always admit the solution $\rho=0$ upon which the theory reverts to its original seed dynamics. In this sense the entire physical content of the disformed theory including the regular branch is represented in the singular branch of solutions. 

Crucially, equation of motion for $\rho$ given by \eqref{eq:div rho universal}, the Einstein equation \eqref{eq:Enstein_low_index} and the Hamilton-Jacobi equation \eqref{eq:EoM_for_phi_in_g} can be expressed purely using the metric $g_{\mu\nu}$, which we will call the {\it physical metric}. 

\section{Examples}\label{Examples}
Let us illustrate our finding from previous sections on some simple examples. We chose these examples so that they do not satisfy the condition \eqref{eq:regularity condition natalie}. Hence, as it was believed before this work, the corresponding disformed theories would not have any new degrees of freedom.
\subsection*{Simple Conformal Transformation}\label{Example_Con}
We first consider a simple, purely conformal transformation
\begin{equation}
    g_{\mu\nu}=e^{Y-1}h_{\mu\nu}\ ,\label{ex:conformal transformation action}
\end{equation}
applied to the Einstein-Hilbert action. The $h_{\mu\nu}$ – equation of motion \eqref{eq:general disformal EoM} is
\begin{equation}
    e^{Y-1}\left (G^{\mu\nu}-G^{\rho\sigma}h_{\rho\sigma}\phi^{,\mu}\phi^{,\nu}\right )=0\ ,\label{ex:conformal transformation EoM}
\end{equation}
where the Einstein tensor as well as its trace is calculated purely through the physical metric $g_{\mu\nu}$. The eigenvector $\xi^{\star}_{\mu\nu}$ given by \eqref{eq:eigenvectors zuma} for the above conformal transformation is just
\begin{equation}
    \xi^{\star}_{\mu\nu}=g_{\mu\nu}\ .
\end{equation}
Hence, contracting equation \eqref{ex:conformal transformation EoM} with this eigenvector gives
\begin{equation}
    G\,e^{Y-1}\left (1-Y\right )=0\ .
\end{equation}
This equation clearly has two branches: either $G$ is equal to zero, which, after plugging it back in the above equation, clearly corresponds to GR in a vacuum. Or there is the second branch, which is characterized by the vanishing of the second factor:
\begin{equation}
    e^{Y-1}\left (1-Y\right )=0\ .
\end{equation}
This is exactly the singularity condition \eqref{eq:singularity condition}, whose vanishing is necessary for the appearance of the novel dynamics. In this particular case the above equation has a single solution $Y_{\star}=1$, which can be plugged back into our equations to give
\begin{equation}
    G_{\mu\nu}-G\phi_{,\mu}\phi_{,\nu}=0\ ,
    \label{eq:Einstein_Ex_1}
\end{equation}
which can be directly obtained from general equation \eqref{eq:Enstein_low_index}. Here every term is evaluated with respect to the physical metric $g_{\mu\nu}$. These equations are clearly equivalent to those of mimetic dark matter. Furthermore, the solution $Y_{\star}=1$ implies that $X_{\star}=1$, which can be calculated from equation \eqref{eq:X(Y)}. Due to this constraint on $X$, the trace of \eqref{eq:Einstein_Ex_1} becomes an identity. Thus, there is one less equation in comparison to the standard GR. The missing information is instead carried by equation of motion \eqref{eq:continuity} for $\phi$ which reads
\begin{equation}
    \nabla_{\mu}\left (G\nabla^{\mu}\phi\right )=0\ .
\end{equation}
This equation is also evaluated purely using the metric $g_{\mu\nu}$. The above set of equations clearly provides us with a solution for $g_{\mu\nu}$ rather then a solution for $h_{\mu\nu}$. However, the corresponding solution for $h_{\mu\nu}$ can be determined due to the constraint $Y_{\star}=1$. Indeed, plugging this into the original transformation \eqref{eq:disformal transformation} we get
\begin{equation}
    g_{\mu\nu}=h_{\mu\nu}\ .
\end{equation}
Note that this is a significant difference in comparison to the original formulation of mimetic dark matter where $Y$ is a pure gauge quantity and thus the metric $h_{\mu\nu}$ cannot be determined uniquely.

\subsection*{Simple Disformal Transformation}\label{Example_Dis}
Let us now examine an example where we get solutions $\phi_{\star}(x)$ with both spacelike as well as timelike gradients. We consider the following disformal transformation
\begin{equation}
    g_{\mu\nu}=h_{\mu\nu}+Y\phi_{,\mu}\phi_{,\nu}\ ,\label{ex:disformal transformation}
\end{equation}
and the Einstein-Hilbert action as the seed theory. The  $h_{\mu\nu}$ – equation of motion \eqref{eq:general disformal EoM} is 
\begin{equation}
    G^{\mu\nu}-G^{\sigma\rho}\,\phi_{,\sigma}\phi_{,\rho}\,\phi^{,\mu}\phi^{,\nu}=0\ .\label{ex:disformal transformation EoM}
\end{equation}
The eigenvector $\xi^{\star}_{\mu\nu}$ given by \eqref{eq:eigenvectors zuma} is
\begin{equation}
    \xi^{\star}_{\mu\nu}=\phi_{,\mu}\phi_{,\nu}\ .
\end{equation}
Contracting equation \eqref{ex:disformal transformation EoM} with this eigenvector $\xi^{\star}_{\mu\nu}$ we obtain
\begin{equation}
    G^{\mu\nu}\,\phi_{,\mu}\phi_{,\nu}\left (1-Y^{2}\right )=0\ .
\end{equation}
This again leads to two branches of solutions: the GR branch (or regular branch) characterized by $G^{\mu\nu}\partial_{\mu}\phi\partial_{\nu}\phi=0$ upon which we can retrieve the Einstein vacuum equations from \eqref{ex:disformal transformation EoM}, and the new branch characterized by
\begin{equation}
    1-Y^{2}=0\ .
\end{equation}
This is again singularity condition \eqref{eq:singularity condition}, which is satisfied by $Y^{\pm}_{\star}=\pm 1$. Via \eqref{eq:X(Y)} these solutions correspond to $X^{\pm}_{\star}=\pm 1/2$. Using these solutions one can lower the indices with metric $g_{\mu\nu}$ to rewrite \eqref{ex:disformal transformation EoM} as
\begin{equation}
    G_{\mu\nu}-4G^{\sigma\rho}\,\phi_{,\sigma}\phi_{,\rho}\,\phi_{,\mu}\phi_{,\nu}=0\ .\label{ex:disformal transformation EoM 2}
\end{equation}
It is not too hard to see that this equation is equivalent to the mimetic dark matter. Indeed, by taking the trace we find:
\begin{equation}
    G=\pm2G^{\sigma\rho}\,\phi_{,\sigma}\phi_{,\rho}\ ,
\end{equation}
which can be plugged back into \eqref{ex:disformal transformation EoM 2} to obtain
\begin{equation}
    G_{\mu\nu}=\pm2G\,\,\phi_{,\mu}\phi_{,\nu}\ ,
\end{equation}
which can be directly obtained from general equation \eqref{eq:Enstein_low_index}. Here the sign corresponds to the chosen branch $Y^{+}_{\star}$ or $Y^{-}_{\star}$. Equation of motion \eqref{eq:div rho universal} reads 
\begin{equation}
    \nabla_{\mu}\left[\rho\,\nabla^{\mu}\phi \right]=0\ .
\end{equation}
This example illustrates that in general there are multiple solutions of the equation $\lambda_{\star}=0$. In this example one is timelike while the other is spacelike. Note that the above equations only describe irrotational dust for the timelike solution, $Y^{+}_{\star}$. The spacelike branch does not have such simple intuitive description, though some discussions have already been done, see e.g. \cite{Deruelle:2014zza,Gorji:2020ten}. Interestingly, physics described by the first purely conformal example and the branch $Y^{+}_{\star}$ of the current disformal example is identical. Thus, different disformal  transformations can result in the same physics, at least for particular branches. 
\subsection*{To Solve or not to Solve}\label{Solve_or_not}
To argue that most functions $C\left(Y,\phi\right)$ and $D\left(Y,\phi\right)$ would allow for solution of equation \eqref{eq:singularity condition} let us consider these functions to be analytical around $Y=0$. Up to second order we have 
\begin{align}
    C\left(Y,\phi\right)&=c_{0}\left(\phi\right)+c_{1}\left(\phi\right)Y+c_{2}\left(\phi\right)Y^{2}\ ,\nonumber\\
    D\left(Y,\phi\right)&=d_{0}\left(\phi\right)+d_{1}\left(\phi\right)Y+d_{2}\left(\phi\right)Y^{2}\ .\label{eq:analytic}
\end{align}
Taking into account that $Y\neq0$, existence of solutions for \eqref{eq:singularity condition} requires to reach equality between 
\begin{equation}
\frac{c_{0}\left(\phi\right)}{Y^{2}}-c_{2}\left(\phi\right)\,,\quad\text{and}\quad d_{1}\left(\phi\right)+2d_{2}\left(\phi\right)Y\ .
\end{equation}
Clearly, for non-vanishing $c_{0}\left(\phi\right)$ and $d_{2}\left(\phi\right)$ there is always at least one solution $Y_{\star}(\phi)$. The maximal possible number of branches is three, with two branches which are of the same sign. Note that $c_{1}\left(\phi\right)$ and $d_{0}\left(\phi\right)$ are irrelevant for the existence of solutions $Y_{\star}(\phi)$. Thus, only particularly constructed coefficient functions $c_{i}\left(\phi\right)$ and $d_{i}\left(\phi\right)$ would allow for the absence of the additional degree of freedom in the disformed theory. Thus, emergence of mimetic DM is a typical result of general disformal transformations.

\section{Mimetic with and without the Weyl Invariance}\label{Mimetic Weyl}
Following the discussion in \cite{Hammer:2015pcx} we may introduce a new set of variables, which conveniently isolates the novel degrees of freedom, which are responsible for the appearance of mimetic DM dynamics in the disformed theory. For simplicity, let us first discuss a simplified case where the disformal transformation is purely conformal, so that
\begin{equation}
    g_{\mu\nu}=C(Y,\phi)h_{\mu\nu}\ .\label{eq:conformal transformation}
\end{equation}
We will revisit the full disformal case in the next section. For our seed theory we consider a general action
\begin{equation}
    S_{\text{seed}}[g,\Psi_M]=\int d^{4}x\,\mathcal{L}[g,\Psi_M]\ .\label{eq:general theory}
\end{equation}
In the following we will not indicate any explicit dependence on matter fields $\Psi_M$ as they do not play any role in the discussion. This dependence can be easily recovered if needed. Moreover, we included the measure $\sqrt{-g}$ into Lagrangian density $\mathcal{L}(g)$. We first introduce a Lagrange constraint and a novel scalar field $\chi$
\begin{equation}
    S_{\text{dis}}[h,\phi,\Lambda,\chi]=\int d^{4}x\left [\,\mathcal{L}(g(h,\phi))+\sqrt{-h}\,\Lambda\left (C(Y,\phi)-\chi\right )\right ]\ .
\end{equation}
The addition of such constraint has no effect on the theory because any changes to the equations of motion are proportional to $\Lambda$, which is constrained to be zero by the $\chi$ equation of motion. However, the introduction of this constraint allows us to substitute $C(Y,\phi)=\chi$ anywhere in the action outside of the constraint itself \cite{Pons:2009ch} without changing the theory. In this way we can use it to eliminate the dependence on $C(Y,\phi)$ in the physical metric $g_{\mu\nu}$:
\begin{equation}
    g_{\mu\nu}=Ch_{\mu\nu}\quad\rightarrow\quad g_{\mu\nu}=\chi h_{\mu\nu}\ .\label{eq:eliminating C}
\end{equation}
By doing so the action becomes
\begin{equation}
    S_{\text{dis}}[h,\phi,\Lambda,\chi]=\int d^{4}x\left [\,\mathcal{L}(\chi h_{\mu\nu})+\sqrt{-h}\,\Lambda\left (C(Y,\phi)-\chi\right )\right ]\ .
\end{equation}
Note that after these steps the $\chi$ equation of motion no longer forces $\Lambda$ to vanish. Instead, the non-trivial value of $\Lambda$ ends up compensating for the changes to the equations of motion which appear due to substitution \eqref{eq:eliminating C}. Crucially, we may now introduce the novel variables\footnote{
%
%
 Effects of field redefinition \eqref{eq:field redefinition conformal} on the dynamics of the theory are determined by the associated Jacobian matrix. Comparing the Jacobian for a conformal transformation \eqref{eq:conformal transformation} with the Jacobian of field redefinition \eqref{eq:field redefinition conformal} it becomes very clear why the two operations are so different. Indeed for the latter the Jacobian is purely diagonal
\begin{equation}
    \mathcal{J}^{\sigma\rho}_{\mu\nu}=\frac{\delta \tilde{g}_{\mu\nu}}{\delta h_{\mu\nu}}\chi \delta^{\rho}_{\mu}\delta^{\sigma}_{\nu}\ .\label{eq:Jacobian easy}
\end{equation}
Consequently, the matrix has but a single eigenvalue $\chi$, which corresponds to $\lambda_{0}$ from previous sections.}
\begin{align*}
    \tilde{g}_{\mu\nu}&=\chi\, h_{\mu\nu}\ ,\\
    \tilde{\Lambda}&=\chi^{-1}\Lambda\ ,\\
    \chi&=\chi\ ,\\
    \phi&=\phi\ . \numberthis{} \label{eq:field redefinition conformal}
\end{align*}
Note that an additional condition $\chi\neq 0$ must be imposed by hand, as the redefinition would be ill-defined for such cases. In the new variables the disformed action reduces to
\begin{equation}
    S_{\text{dis}}[\tilde{g},\phi,\tilde{\Lambda},\chi]=\int d^{4}x\left [\,\mathcal{L}(\tilde{g})+\sqrt{-\tilde{g}}\,\tilde{\Lambda}\,\left (\chi^{-1}C(\chi\tilde{X},\phi)-1\right )\right ]\ ,\label{eq:action constraint conformal}
\end{equation}
where $\tilde{X}=\tilde{g}^{\mu\nu}\partial_{\mu}\phi\partial_{\nu}\phi$. We can see that the above procedure allowed us to isolate the effect of the conformal transformation \eqref{eq:conformal transformation}, which effectively boiled down to the introduction of an extra Lagrange constraint with a very specific dependence on the scalar field $\chi$.

The key difference between the general setup \eqref{eq:conformal transformation} and the specific mimetic dark matter case with $C(Y,\phi)=b(\phi)Y$, where $b(\phi)$ is an arbitrary function of $\phi$, is that in the latter case $\chi$ drops out of the action entirely. This reflects the underlying Weyl invariance of the mimetic DM setup. For any theory disformed via \eqref{eq:conformal transformation}, the Weyl invariance is broken and the field dependence on $\chi$ remains in the action. Crucially, this dependence is purely algebraic, there are no derivatives of $\chi$ in action \eqref{eq:action constraint conformal}. This means that $\chi$ is an auxiliary field which can be integrated out of the action. 
The equation of motion for $\chi$ reads
\begin{equation}
    \tilde{\Lambda}\Big [\chi^{-1}C_{Y}(\chi \tilde{X},\phi)\tilde{X}-\chi^{-2}C(\chi \tilde{X},\phi)\Big]=0\ .\label{eq:EoM chi conformal full}
\end{equation}
This is satisfied when
\begin{equation}
    C_{Y}(\chi \tilde{X},\phi)\tilde{X}\chi=C(\chi \tilde{X},\phi)\ .\label{eq:EoM chi conformal}
\end{equation}
It is easy to recognise in the above relation the singularity condition given by \eqref{eq:singularity condition}. Note that in general the $\chi$ equation of motion also admits the solution $\tilde{\Lambda}=0$, which corresponds to the solutions of the original seed theory \eqref{eq:general theory}. A key feature of equation \eqref{eq:EoM chi conformal} above is that $\chi$ enters only in combination with $\tilde{X}$ as $\chi\tilde{X}$. Thus, \eqref{eq:EoM chi conformal} effectively depends only on two variables $\chi\tilde{X}$ and $\phi$.
Therefore, any solution of this algebraic equation with respect to $\chi\tilde{X}$ (if such solution exists) has the form
\begin{equation}
    \chi_{\star} \tilde{X}=Y_{\star}(\phi)\ ,\label{eq:implicit equation solution}
\end{equation}
where we used the same notation for the roots of singularity equation as in section \ref{Jacobi}. 
Plugging this particular branch $\chi_{\star}$ back into action \eqref{eq:action constraint conformal} yields
\begin{equation}
    S^{\star}_{\text{dis}}[\tilde{g},\phi,\tilde{\Lambda}]=\int d^{4}x\left [\,\mathcal{L}(\tilde{g})+\sqrt{-\tilde{g}}\,\tilde{\Lambda}\,\left (\frac{\tilde{X}}{X_{\star}(\phi)}-1\right )\right ]\ ,\label{eq:action weyl restored conformal}
\end{equation}
where we substituted $Y_{\star}(\phi)$ into equation \eqref{eq:X(Y)}. 
This is clearly a reduced form of the action for mimetic DM given by the mimetic ansatz
\begin{equation}
    g_{\mu\nu}=\left[\frac{h^{\alpha\beta}\phi_{,\alpha}\phi_{\beta}}{X_{\star}\left(\phi\right)}\right]\,h_{\mu\nu}\,.
    \label{eq:final_ansatz}
\end{equation}
and thus corresponds to a manifestly Weyl invariant theory. This expression is identical to the original mimetic ansatz from \cite{Chamseddine:2013kea} if one uses redefined scalar field $\Phi$ given by equation \eqref{eq:PHI}. Contrary to action \eqref{eq:action weyl restored conformal}, which is reduced to just one solution $Y_{\star}(\phi)$ of the singularity equation, action \eqref{eq:action constraint conformal} preserves information about all branches of solutions for the singularity equation.

\section{Equivalent Action for Theory with Singular Disformal Transformation}\label{Action_Disformal}
In the case of the fully-fledged disformal transformation \eqref{eq:disformal transformation}
we need to isolate two free functions $C$ and $D$. We use the same strategy as before, which now requires to introduce two Lagrange constraints instead of one:
\begin{equation}
    S_{\text{dis}}[h,\phi,\Lambda,\alpha,\beta]=\int d^{4}x\,\Bigg[\,\mathcal{L}(g(h,\phi))+\Lambda_{\alpha}\sqrt{-h}\big (C(Y,\phi)-\alpha\big )+\Lambda_{\beta}\sqrt{-h}\big (D(Y,\phi)-\beta\big )\Bigg ]\ .\label{eq:action full disformal constraint}
\end{equation}
This allows us to eliminate the dependence on $C(Y,\phi)$ and $D(Y,\phi)$ in the action outside of the constraints. Furthermore, using the scalars $\alpha$ and $\beta$, we are able to propose a new metric variable analogical to \eqref{eq:field redefinition conformal} from the conformal case
\begin{equation}
    \bar{g}_{\mu\nu}=\alpha\, h_{\mu\nu}+\beta\,\partial_{\mu}\phi\partial_{\nu}\phi\ .\label{eq:metric field redefinition}
\end{equation}
Again, a field redefinition like this is, in principle, able to produce novel dynamics; however, this does not happen here because $\alpha$ and $\beta$ here are free fields that do not depend on the metric $h_{\mu\nu}$ in any way. Consequently the Jacobian of the above mapping has a very simple form \eqref{eq:Jacobian easy}. In order to rewrite the action \eqref{eq:action full disformal constraint} in terms of the new variable $\bar{g}_{\mu\nu}$ it is useful to derive the relation between various quantities defined with respect to the two metrics $h_{\mu\nu}$ and $\bar{g}_{\mu\nu}$. Assuming $\alpha\neq 0$ the mapping \eqref{eq:metric field redefinition} can be inverted to give
\begin{equation}
    h_{\mu\nu}=\alpha^{-1}\Big [\Tilde{g}_{\mu\nu}-\beta\,\partial_{\mu}\phi\partial_{\nu}\phi\Big ]\ .
\end{equation}
The inverse metric $h^{\mu\nu}$ is then
\begin{equation}
    h^{\mu\nu}=\alpha\Big [\bar{g}^{\mu\nu}+\frac{\beta}{1-\beta\bar{X}}\,\bar{\partial}^{\mu}\phi\bar{\partial}^{\nu}\phi\Big ]\,.
\end{equation}
Here $\bar{g}^{\mu\nu}$ is inverse of $\bar{g}_{\mu\nu}$, and the bar above the partial derivatives denotes that their index has been raised using the contravariant metric $\bar{g}^{\mu\nu}$. We also define the associated kinetic term $\bar{X}=\bar{g}^{\mu\nu}\partial_{\mu}\phi\partial_{\nu}\phi$. Contracting both sides with $\partial_{\mu}\phi\partial_{\nu}\phi$ we find the relation between $\bar{X}$ and $Y$ to be
\begin{equation}
    Y=\frac{\alpha\bar{X}}{1-\beta\bar{X}}\ .
\end{equation}
Finally, the determinant of $h_{\mu\nu}$ is related (see e.g. \cite{Armendariz-Picon:2005oog}) to the determinant of $\bar{g}_{\mu\nu}$ as
\begin{equation}
    \det h_{\mu\nu}=\alpha^{-4}\det \bar{g}_{\mu\nu}\big (1-\beta\bar{X})\ .
\end{equation}
We can see that in order for the above relations to be well defined we need to pose additional requirements on the scalar fields $\alpha$ and $\beta$. This stems from the fact that even when $h_{\mu\nu}$ is a proper metric tensor and the mapping \eqref{eq:metric field redefinition} is regular and invertible, it is not guaranteed that $\bar{g}_{\mu\nu}$ has all the characteristics of a metric tensor and vice versa. These properties need to be ensured by hand, as discussed for the disformal transformations in \cite{Bekenstein:1992pj,Bettoni:2013diz,Bruneton:2006gf}. This translates into $\alpha\neq 0$ and $\beta\bar{X}< 1$, which enable the invertibility of $\bar{g}_{\mu\nu}$ and also ensure that $\bar{g}_{\mu\nu}$ has a proper Lorentzian signature. Substituting the above relations into action \eqref{eq:action full disformal constraint} we obtain
\begin{align*}
    S_{\text{dis}}[\bar{g},\phi,\Lambda,\alpha,\beta]=\int d^{4}x\,\Bigg[\,\mathcal{L}(\bar{g})+\sqrt{-\bar{g}}\,\alpha^{-2}\sqrt{1-\beta\bar{X}}\left(\Lambda_{\alpha}\,\psi_{\alpha}+\Lambda_{\beta}\,\psi_{\beta}\right) \Bigg ]\ ,\numberthis
\end{align*}
where we have introduced
\begin{align*}
    \psi_{\alpha}&=C\Big(\frac{\alpha\bar{X}}{1-\beta\bar{X}},\phi\Big)-\alpha\ ,\\
    \psi_{\beta}&=D\Big(\frac{\alpha\bar{X}}{1-\beta\bar{X}},\phi\Big)-\beta\ .
\end{align*}
Due to our assumption $\beta\bar{X}\neq 1$ the factors following the Lagrange multipliers $\Lambda_{\alpha}$ and $\Lambda_{\beta}$ are strictly non-vanishing and therefore we are allowed to absorb them by the following redefinition
\begin{align*}
    \bar{\Lambda}_{\alpha}&=\Lambda_{\alpha}\,\alpha^{-2}\sqrt{1-\beta\bar{X}}\ ,\\
    \bar{\Lambda}_{\beta}&=\Lambda_{\beta}\,\alpha^{-2}\sqrt{1-\beta\bar{X}}\ .\numberthis
\end{align*}
Hence, the action simplifies to
\begin{equation}
    S_{\text{dis}}[\bar{g},\phi,\bar{\Lambda},\alpha,\beta]=\int d^{4}x\,\Bigg[\mathcal{L}(\bar{g})+\sqrt{-\bar{g}}\,\bar{\Lambda}_{\alpha}\psi_{\alpha}+\sqrt{-\bar{g}}\,\bar{\Lambda}_{\beta}\psi_{\beta}\Bigg ]\ .\label{eq:two constraint action}
\end{equation}
Due to the redefinition $h_{\mu\nu}\rightarrow\bar{g}_{\mu\nu}$ the Lagrange multipliers $\alpha$ and $\beta$ became auxiliary fields. However, their equations of motion are not as simple as in the conformal case from the previous section. The main difficulty stems from the presence of two Lagrange multipliers in their equations of motion $\bar{\Lambda}_{\alpha}$ and $\bar{\Lambda}_{\beta}$. Indeed the solution for either $\alpha$ or $\beta$ from their own equations of motion will in general depend on $\bar{\Lambda}_{\alpha}$ or $\bar{\Lambda}_{\beta}$, which is qualitatively very different from what we encountered in the conformal case. Interestingly, we have found that these difficulties can be avoided by introducing an additional auxiliary variable $\theta$, which allows us to reduce the number of constraints in the action to just one. The variable $\theta$ is defined as
\begin{equation}
    \theta=\frac{\alpha}{1-\beta\bar{X}}\ ,
\end{equation}
which we enforce through yet another Lagrange constraint
\begin{equation}
    \sqrt{-\bar{g}}\,\Lambda_{\theta}\Bigg(\frac{\alpha}{1-\beta\bar{X}}-\theta\Bigg)\ .
\end{equation}
This term is meant to be added to action \eqref{eq:two constraint action}. We can now substitute this constraints to the two original constraints $\psi_{\alpha,\beta}$ to obtain
\begin{align}
    \psi_{\alpha}=&C\Big(\theta \bar{X},\phi\Big)-\alpha\ ,\\
    \psi_{\beta}=&D\Big(\theta \bar{X},\phi\Big)-\beta\ .
\end{align}
Then we can use these constraints to eliminate $\alpha$ and $\beta$ from the $\theta$-constraint, which becomes
\begin{equation}
    \Lambda_{\theta}\Bigg(\frac{C(\theta \bar{X},\phi)}{1-D(\theta \bar{X},\phi)\bar{X}}-\theta\Bigg)\ .
\end{equation}
This way $\alpha$ and $\beta$ only appear linearly in their respective constraints and therefore they are simple Lagrange multipliers. Hence, they can be integrated out of the action. This means that we may remove the constraints $\psi_{\alpha,\beta}$ without changing the dynamical content of the theory, so that the action becomes
\begin{equation}
    S_{\text{dis}}[\bar{g},\phi,\Lambda_{\theta},\theta]=\int d^{4}x\,\Bigg[\mathcal{L}(\bar{g})+\sqrt{-\bar{g}}\,\Lambda_{\theta}\Bigg(\frac{C(\theta \bar{X},\phi)}{1-D(\theta \bar{X},\phi)\bar{X}}-\theta\Bigg)\Bigg ]\ .\label{eq:constraint action proto}
\end{equation}
The equation of motion for $\theta$ obtained from this action does not have the same properties as equation \eqref{eq:EoM chi conformal}. Namely $\theta$ does not appear in the combination $\bar{X}\theta$ everywhere as $\chi$ did. However, the terms where this property fails are proportional to the above constraint itself. Consequently, this shortcoming can be remedied by a field redefinition of the Lagrange multiplier $\Lambda_{\theta}$ itself
\begin{equation}
    \bar{\Lambda}_{\theta}=\frac{\theta\Lambda_{\theta}}{1-D(\theta \bar{X},\phi)\bar{X}}\ .
\end{equation}
This finally leaves us with the action 
\begin{equation}
    S_{\text{dis}}[\bar{g},\phi,\bar{\Lambda}_{\theta},\theta]=\int d^{4}x\,\Bigg[\mathcal{L}(\bar{g})+\sqrt{-\bar{g}}\,\bar{\Lambda}_{\theta}\Bigg(\theta^{-1}C(\theta \bar{X},\phi)+D(\theta \bar{X},\phi)\bar{X}-1\Bigg)\Bigg ]\ .\label{notes:one constraint action}
\end{equation}
It is important to stress that by deriving this action we have effectively isolated the entire effect of the disformal transformation into a single constraint. Indeed, the change in the action of the seed theory is reduced to mere replacement $g_{\mu\nu}\rightarrow\bar{g}_{\mu\nu}$. The new constraint depends on an extra auxiliary field $\theta$, which is in general non-trivial and yields its own equation of motion\footnote{
Note that in the case of special disformal transformations \eqref{eq:regularity condition natalie} the $\theta$ dependence in the above constraint drops out entirely, just like it has in the case of the mimetic dark matter \cite{Hammer:2015pcx}. Indeed, condition \eqref{eq:regularity condition natalie} applied to the above combination of terms gives
\begin{equation}
    \theta^{-1}\left (C(\theta \bar{X},\phi)+D(\theta \bar{X},\phi)\theta\bar{X}\right )=c(\phi) \bar{X}\ .
\end{equation}
 
}
\begin{equation}
    \theta^{-2}\bar{\Lambda}_{\theta}\Big (C(\theta \bar{X},\phi)-C_{Y}(\theta \bar{X},\phi)\theta \bar{X}-D_{Y}(\theta \bar{X},\phi)\theta^{2}\bar{X}^{2}\Big )=0\ .\label{eq:theta EoM with Lambda}
\end{equation}
This equation is qualitatively the same as equation \eqref{eq:EoM chi conformal full}. In particular, it provides two possible branches of solutions, which are associated with the vanishing of either $\bar{\Lambda}_{\theta}$ or
\begin{equation}
    C(\theta \bar{X},\phi)-C_{Y}(\theta \bar{X},\phi)\theta \bar{X}-D_{Y}(\theta \bar{X},\phi)\theta^{2}\bar{X}^{2}=0\ .\label{eq:theta equation auxiliary}
\end{equation}
Again, it is easy to recognise in the above relation the singularity condition given by \eqref{eq:singularity condition}, where $Y=\theta \bar{X}$. The first option, vanishing $\bar{\Lambda}_{\theta}$, corresponds to the regular branch of the transformed theory, which is equivalent to the solutions of the original seed theory. Indeed, any change to the equations of motion of the theory \eqref{notes:one constraint action} is necessarily proportional to $\Lambda_{\theta}$. Hence, upon $\bar{\Lambda}_{\theta}=0$ the equations of motion reduce to those of the original seed theory. In particular, if singularity condition \eqref{eq:singularity condition} does not have any solutions, equation of motion \eqref{eq:theta EoM with Lambda} enforces $\bar{\Lambda}_{\theta}=0$ and makes the disformed theory to be equivalent to the seed one. In this regular case the variation with respect to $\bar{\Lambda}_{\theta}$ yields constraint $\psi_{\theta}=C+DY-\theta=0$. In accordance with the implicit function theorem, this constraint can be solved with respect to $\theta$, because  $\partial\psi_{\theta}/\partial\theta=-\lambda_{\star}/\theta\neq0$.

The dependence on $\theta$ in \eqref{eq:theta equation auxiliary} enters strictly in combination $\theta\bar{X}$. Therefore, in the singular case, any solution of this equation must be of the form
\begin{equation}
    \theta_{\star}=\bar{X}^{-1}Y_{\star}(\phi)\ ,
\end{equation}
where we again used the same notation for the roots of singularity equation as in section \ref{Jacobi}. Since we have found a solution for $\theta$ from its own equation of motion we can substitute it back into the action \eqref{notes:one constraint action}. By doing so the constraint becomes that of mimetic dark matter so that
\begin{equation}
    S^{\star}_{\text{dis}}[\bar{g},\phi,\bar{\Lambda}_{\theta}]=\int d^{4}x\,\Bigg[\mathcal{L}(\bar{g})+\sqrt{-\bar{g}}\,\bar{\Lambda}_{\theta}\,\left (\frac{\bar{X}}{X_{\star}(\phi)}-1\right )\Bigg ]\ ,
\end{equation}
where we again substituted $Y_{\star}(\phi)$ into equation \eqref{eq:X(Y)}. Thus, similarly to the case of conformal transformation, the disformed action reduced to the one of mimetic gravity with the same mimetic ansatz \eqref{eq:final_ansatz}. Thus, the dynamics of the disformed theory is described by the Weyl-invariant action.

Finally, comparing \eqref{notes:one constraint action} with \eqref{eq:action constraint conformal}, one infers that for any disformal transformation given by functions $C\left(Y,\phi\right)$ and $D\left(Y,\phi\right)$ one can always find a dual, purely conformal, transformation with the conformal factor $\widetilde{C}\left(Y,\phi\right)=C\left(Y,\phi\right)+D\left(Y,\phi\right)Y$. The dynamics of both transformed theories is the same, including potential branches of solutions of singularity condition \eqref{eq:singularity condition}.


\section{Conclusions and discussion}\label{Conclusions}
We derived and discussed novel consequences of applying a generic disformal transformation \cite{Bekenstein:1992pj} of the metric  
\begin{equation}
g_{\mu\nu}=C\left(Y,\phi\right)h_{\mu\nu}+D\left(Y,\phi\right)\partial_{\mu}\phi\,\partial_{\nu}\phi\ ,\label{eq:conc:disformal transformation}
\end{equation}
where 
\begin{equation}
Y=h^{\mu\nu}\partial_{\mu}\phi\,\partial_{\nu}\phi\,,\label{eq:concY}
\end{equation}
simultaneously\footnote{Obviously, our analysis is also directly applicable to purely conformal transformations, $D=0$, performed only in the gravity sector, provided the matter sector is Weyl-invariant.} in gravity and matter sectors. These sectors are considered to be originally independent of the scalar field $\phi$.
For the functions $C\left(Y,\phi\right)$ and $D\left(Y,\phi\right)$ related as  
\begin{equation}
    D(Y,\phi)=-\frac{C(Y,\phi)}{Y}+c(\phi)\ ,\label{eq:conc:regularity condition natalie}
\end{equation}
it has been known \cite{Deruelle:2014zza} that the transformation results in the appearance an extra degree of freedom. As it was demonstrated there, this degree of freedom either corresponds to the mimetic dark matter \cite{Chamseddine:2013kea}, or to its analytic continuation to spacelike gradients of $\phi$. The novel finding of our work is that this emergence of the new degree of freedom is not limited to this special choice of functions defining the transformation.  Instead, we have found that the novel degree of freedom is introduced whenever the weaker condition
\begin{equation}
    C=C_{Y}Y+D_{Y}Y^{2}\ ,\label{eq:conc:regularity condition zuma}
\end{equation}
is satisfied as a differential equation for $\phi$. Namely, implicit function theorem provides algebraic solution  $Y_\star(\phi)$ of this equation, then equation of motion for the scalar field  
\begin{equation}
h^{\mu\nu}\partial_{\mu}\phi\,\partial_{\nu}\phi=Y_\star(\phi)\,,
\label{eq:concl_Hamilton_Jacobi}
\end{equation}
is of the type of the relativistic Hamilton-Jacobi equation \cite{Landafshitz_Teorpol}. Clearly, one can always select such functions $C\left(Y,\phi\right)$ and $D\left(Y,\phi\right)$ that condition \eqref{eq:conc:regularity condition zuma} is never satisfied\footnote{Alternatively one could admit existence of solutions that are physically not-viable such as solutions satisfying $C=0$.} for real valued field $\phi$. In that case, the transformation is invertible and is "safe" - does not add new unexpected degrees of freedom. However, as we argued in this work, generic functions $C\left(Y,\phi\right)$ and $D\left(Y,\phi\right)$ would allow for some real solutions $Y_\star(\phi)$ of \eqref{eq:conc:regularity condition zuma}. Thus, generic disformal transformation is not "safe" with respect to adding new unexpected degree of freedom.
We have shown that the emergent degree of freedom again corresponds to mimetic gravity \cite{Chamseddine:2013kea}. Here our main results are: scalar equation of motion \eqref{eq:div rho universal} and tensor equation of motion  \eqref{eq:Enstein_low_index}. Importantly, this happens in a much wider class of metric transformations than previously thought, see e.g. \cite{Deruelle:2014zza,Arroja:2015wpa,Takahashi:2017pje}. We argued that actually most disformal transformation will result in such an additional degree of freedom, see section \ref{Examples}. 




We have traced the origin of the extra degrees of freedom to the vanishing of the Jacobian determinant of the disformal transformation. Indeed, the condition \eqref{eq:conc:regularity condition zuma} is equivalent to the vanishing of a particular eigenvalue of the Jacobian matrix \eqref{eq:Jacobian matrix}, calculated in  \cite{Zumalacarregui:2013pma}. We have shown that the eigenvectors associated to this eigenvalue play a distinguished role in our construction and they have a clear physical interpretation. Namely, in the dynamical regime with the extra degree of freedom, eigenvector \eqref{eq:eigenvectors zuma} generates a local symmetry of the theory  while \eqref{eq:eigenvectors dual} is proportional to the energy-momentum tensor of the induced mimetic dark matter (or its analytical continuation). Since these vectors always come as a pair, we may view the mimetic dark matter as a consequence of this emergent local symmetry of the theory.

The resulting matter sector is largely equivalent to the addition of mimetic gravity\footnote{In particular, compact objects would be described in the same way as in mimetic gravity, see e.g. \cite{Gorji:2020ten}.}. The only subtle difference is that, various choices of the functions $C$ and $D$ may lead to multiple branches $Y^i_{\star}(\phi)$ with distinct solutions for the field $\phi_{\star}(x)$. The choice of the branch represents additional discrete freedom in the initial data. 
It is well known that mimetic DM corresponds to fluid-like irrotational dust \cite{Golovnev:2013jxa,Barvinsky:2013mea,Hammer:2015pcx}. The latter is plagued\footnote{For possible exceptions see \cite{Mukohyama:2009tp}.} by caustics formation, see e.g. \cite{Babichev:2016jzg}. It is interesting to investigate whether transition between different branches can help to alleviate the problem. Moreover, fluid-like irrotational dust also appears in the projectable Ho\v{r}ava-Lifshitz gravity \cite{Horava:2009uw,Blas:2009qj,Blas:2009yd,Mukohyama:2009mz} and in some other quantum gravity scenarios, like \cite{Chamseddine:2014nxa,Zlosnik:2018qvg}. It is intriguing to investigate whether one could trace the origin of this dust of quantum gravity to disformal transformations. Interestingly, mimetic gravity extended by higher derivatives may realise conceptually interesting scenario of limiting curvature and model effective dynamics of loop quantum gravity, see e.g. \cite{Chamseddine:2016uef,Chamseddine:2016ktu,Bodendorfer:2017bjt,Langlois:2017hdf,BenAchour:2017ivq,Yoshida:2018kwy,Bodendorfer:2018ptp,Brahma:2018dwx}. One may speculate that the presence of the branches $Y^i_{\star}(\phi)$ may help to solve some problems within quantum gravity, like topological transitions. However, it is also not clear how to treat the presence of the branches in realm of quantum field theories.

On the other hand, mimetic DM allows for a simple UV-completion \cite{Babichev:2017lrx}. Whereas it is not entirely clear whether such a completion exists in the presence of branches $Y^i_{\star}(\phi)$ and solutions with spacelike gradients.


In our argument we have highlighted the role of the Jacobian determinant and of the associated eigenvalues and eigenvectors, without relying on the details of the metric transformation itself. This shows that our results are actually quite general and can be straightforwardly applied to other metric transformations, which do not depend on the metric derivatives. Thus, our procedure can be aptly used to analyse, for example, the vector metric transformations \cite{Papadopoulos:2017xxx,Domenech:2018vqj,DeFelice:2019hxb}. Our results could also be used to show that non-Weyl symmetric generalizations of various instances of mimetic gravity \cite{Hammer:2020dqp,Gorji:2018okn,Gorji:2019ttx} may yield equivalent results. It is interesting to apply similar analysis for disformal transformations in such modified theories of gravity, where the dynamical variables are not limited to metric and scalar fields, see e.g. \cite{Golovnev:2019kcf}.  

Finally, we have shown that the effects of the disformal transformation on any theory can be isolated on the level of the action by the addition of a single Lagrange constraint along with an associated Lagrange multiplier $\Lambda$ and a novel scalar field $\theta$
\begin{equation}
    S_{\mathrm{dis}}[g,\phi,\Lambda,\theta]= S_{\mathrm{seed}}[g,\phi]+\int d^{4}x\,\sqrt{-g}\,\Lambda\Bigg(\theta^{-1}C(\theta X,\phi)+D(\theta X,\phi)X-1\Bigg)\ .\label{eq:conc:one constraint action}
\end{equation}
We have demonstrated that this field $\theta$ is generally an auxiliary field, which can be integrated out of the action. Doing so reduces the action to the original form with the addition of a mimetic constraint. Therefore, the transformed theory is equivalent to a particular mimetic gravity theory, which is Weyl-invariant.

The main obstacle to applying our arguments to metric transformations, which contain derivatives of the metric is that the associated Jacobian \eqref{eq:Jacobian matrix} will no longer be a matrix in the conventional sense. Rather, it will become a distribution as it will contain the derivatives of the Dirac delta function coming from the derivatives of $h_{\mu\nu}$. This in turn implies that equations \eqref{eq:eigenvector equation zuma} and \eqref{eq:eigenvector equation dual} become differential equations. Consequently, the standard notion of eigenvalues and eigenvectors may no longer be applicable. Nevertheless, if such differential equations (\eqref{eq:eigenvector equation zuma} and \eqref{eq:eigenvector equation dual}) have solutions with a vanishing right hand side it would signal novel dynamics, as this would imply additional freedom in the equation \eqref{eq:general disformal EoM}. We expect that the analogue of the field $\rho$ may be subjected to some differential constraint in this case. It would definitely be very interesting to apply such analysis to the derivative extensions of the disformal transformation \cite{Takahashi:2017zgr,Jirousek:2018ago,Takahashi:2021ttd,Babichev:2019twf,Babichev:2021bim}.\par
To conclude, we have demonstrated that disformal transformations of the metric hide a lot of surprises with interesting and unexpected physical consequences.

\section*{Acknowledgments}
Our collaboration is supported by the Bilateral
Czech-Japanese Mobility Plus Project JSPS-21-12 (JPJSBP120212502). The work of P.~J. was supported by the Grant Agency of the Czech Republic (GAČR grant 20-28525S), through the most of the project. P.~J. also acknowledges funding from the South African Research Chairs Initiative of the Department of Science and Technology and the National Research Foundation of South Africa in the final stages of preparation of the manuscript. K.~S. was supported by JSPS KAKENHI Grant Number JP20J12585 for the initial stages of this work. A.~V. acknowledges
support from the European Structural and Investment Funds and the Czech Ministry of Education, Youth and Sports (Project CoGraDS -CZ.02.1.01/0.0/0.0/15003/0000437). M.~Y. acknowledges financial support from JSPS Grant-in-Aid for Scientific Research No. JP18K18764, JP21H01080, JP21H00069. 

\bibliographystyle{utphys}
\bibliography{references} 

\end{document}